\documentclass[twocolumn]{aastex631}
\newcommand{\msun}{${\cal M}_\odot$\,}

\begin{document}

% for float placement:
\renewcommand{\topfraction}{1.0}
\renewcommand{\bottomfraction}{1.0}
\renewcommand{\textfraction}{0.0}

\shorttitle{Speckle Interferometry at SOAR}
\shortauthors{Tokovinin et al.}

\title{Speckle Interferometry at SOAR in 2021 }

\author{Andrei Tokovinin}
\affil{Cerro Tololo Inter-American Observatory | NFSs NOIRLab Casilla 603, La Serena, Chile}
\email{andrei.tokovinin@noirlab.edu}

\author{Brian D. Mason}
\affil{U.S. Naval Observatory, 3450 Massachusetts Ave., Washington, DC, USA}
\email{brian.d.mason.civ@us.navy.mil}
\author{Rene A. Mendez}
\affil{Universidad de Chile,  Casilla 36-D, Santiago, Chile}
\email{rmendez@uchile.cl}
\author{Edgardo Costa}
\affil{Universidad de Chile,  Casilla 36-D, Santiago, Chile}

\begin{abstract}
The  speckle  interferometry  program  at   the  the  4.1  m  Southern
Astrophysical  Research   Telescope  (SOAR),  started  in   2008,  now
accumulated  over 30,300  individual observations  of 12,700  distinct
targets. Its main goal is to monitor orbital motion of close binaries,
including members of high-order hierarchies and low-mass dwarfs in the
solar neighborhood. The results from 2021 are published here, totaling
2,623 measurements of 2,123 resolved  pairs and non-resolutions of 763
targets.  The  median measured separation  is 0\farcs21, and  75 pairs
were closer than 30\,mas. The  calibration of scale and orientation is
based on the observations of  103 wide pairs with well-modeled motion.
These calibrators  are compared to  the latest Gaia data  release, and
minor (0.5\%) systematic errors  were rectified, resulting in accurate
relative positions with  typical errors on the order of  1 mas.  Using
these new  measurements, orbits of   282 binaries  are determined
here (54 first determinations  and 228  corrections).  We
resolved  for the  first time  50 new  pairs, including  subsystems in
known binaries.  A list of 94 likely spurious pairs unresolved at SOAR
(mostly close Hipparcos binaries) is also given.
\end{abstract} 

%---------------------------------------------------------
\section{Introduction}
\label{sec:intro}

This paper  continues the series  of double-star measurements  made at
the 4.1  m Southern Astrophysical  Research Telescope (SOAR) since 2008  with the
speckle camera, HRCam.    Previous results  were published by
\citet[][hereafter           TMH10]{TMH10}            and           in
\citep{SAM09,Hrt2012a,Tok2012a,TMH14,TMH15,SAM15,SAM17,SAM18,SAM19,SAM20}.
Observations reported here were made during 2021.

The structure and content of this  paper are similar to other paper of
this series.  Section~\ref{sec:obs} reviews  all speckle programs that
contributed  to this  paper,  the observing  procedure,  and the  data
reduction.  The results are  presented in Section~\ref{sec:res} in the
form of  electronic tables archived  by the journal.  We  also discuss
new resolutions, present new orbits  resulting from this data set, and
give a list of likely spurious  unresolved pairs.  A short summary and
an outlook of further work in Section~\ref{sec:sum} close the paper.

%---------------------------------------------------------
\section{Observations}
\label{sec:obs}

%-------------------------------------------------------------
\subsection{Observing Programs}

As  in previous  years,  HRCam (see  Section~\ref{sec:inst}) was  used
during 2021  to execute several  observing programs, some  with common
targets.  Table~\ref{tab:programs} gives an overview of these programs
and indicates which  observations are published in  the present paper.
The   numbers  of   observations   are   approximate.  Overall,   5,138
observations were  made during 2021.   Here is a brief  description of
the main programs.

{\it Orbits} of resolved binaries.  New measurements contribute to the
steady  improvement of  the  quantity  and quality  of  orbits in  the
\href{http://www.astro.gsu.edu/wds/orb6.html}{Sixth Catalog  of Visual
  Binary Star Orbits} \citep{VB6}. See \citet{Mendez2022,Gomez2022} as
recent examples  of this  work.  We  provide a  large  table  of updated  and
first-time orbits in Section \ref{sec:orbits}.

{\it Hierarchical  systems} of stars  are of special  interest because
their  architecture is  relevant  to star  formation, while  dynamical
evolution   of  these   hierarchies  increases   chances  of   stellar
interactions  and  mergers  \citep{review}.   Orbital  motions  of
several triple systems  are monitored at SOAR and these  data are used
for the orbit determinations \citep{TL2020,Tok2021a}.
%\citep{MSC}.   

{\it  Hipparcos  binaries} within  200\,pc  are  monitored to  measure
masses of stars  and to test stellar evolutionary  models, as outlined
by, e.g., \citet{Horch2015,Horch2017,Horch2019}.  The southern part of
this  sample is  addressed at  SOAR \citep{Mendez2017}.   This program
overlaps with the general work on visual orbits.

\begin{deluxetable}{ l l l l  } 
\tabletypesize{\scriptsize}    
\tablecaption{Observing programs
\label{tab:programs} }                    
\tablewidth{0pt}     
\tablehead{ \colhead{Program}  &
\colhead{PI}  &  
\colhead{$N$} & 
\colhead{Publ.\tablenotemark{a}} 
}
\startdata
Orbits                & Mason, Tokovinin    & 1092 & Yes \\
Hierarchical systems  & Tokovinin           &  149 & Yes \\
Hipparcos binaries    & Mendez, Horch       &  451 & Yes \\
Neglected binaries    & R.~Gould, Mason     & 624 & Yes \\
Nearby M dwarfs     & E. Vrijmoet           &  323 & No \\
TESS follow-up        & C. Ziegler           & 724  & No \\
Acceleration stars    & K.~Franson          & 388 & No
\enddata
\tablenotetext{a}{This columns indicates whether the results are
  published here (Yes) or deferred to future
  papers (No). }
\end{deluxetable}
% YMG 114

{\it  Neglected  close  binaries}  from  the  Washington  Double  Star
Catalog,      WDS       \citep{WDS},\footnote{See      the      latest
  \href{http://www.astro.gsu.edu}{online} WDS version.}  were observed
as a  `filler' at low priority.  In some cases, we  resolved new inner
subsystems, thus  converting classical visual pairs  into hierarchical
triples.  Some WDS  pairs are  moving fast  near periastron,  allowing
calculation  of  their  first  orbits after  several  observations  at
SOAR. Another result  of this effort is a list  of spurious pairs that
should be removed from the WDS.

{\it Nearby  K and  M dwarfs}  are being observed  at SOAR  since 2018
following the initiative of T.~Henry  and E.~Vrijmoet.  The goal is to
assemble  statistical  data on  orbital  elements,  focusing on  short
periods. The sample includes known  and suspected binaries detected by
astrometric  monitoring, Gaia,  etc.  First  results on  M dwarfs  are
published by  \citet{Vrijmoet2022}. In  2021, we continued  to monitor
these pairs, some of them with fast orbital motion.

{\it TESS follow-up} continues the program executed in 2018--2020. Its
results are published in \citep{TESS,TESS2}.  All speckle observations
of   TESS   targets  of   interest   are   promptly  posted   on   the
\href{https://exofop.ipac.caltech.edu/tess/}{EXOFOP web  site.}  These
data are  used in  the growing  number of  papers on  TESS exoplanets,
mostly as limits on close companions to exohosts.

{\it  Accelerating  stars}  were  observed  as potential  targets  of
high-contrast imaging of exoplanets in a program led by K.~Franson and
B.~Bowler.  A   substantial  fraction  of  nearby   stars  with  small
accelerations,  detected by  comparing Gaia  and Hipparcos  astrometry
\citep{Brandt2021}, are  just binaries with stellar  companions. Their
resolution at SOAR  serves to clean the target  list for high-contrast
imaging of  the remaining stars  (possible exohosts) and  presents an
independent  interest for  future orbit  calculation and  multiplicity
statistics.

Observations of young stars belonging to moving groups were published
in our previous paper \citep{SAM20}. In 2021, a modest number of additional
targets from this program were observed. 

If observations  of a given  star were requested by  several programs,
they are published  here even when the other  program still continues.
We  also  publish here  the  measurements  of previously  known  pairs
resolved during surveys, for example in the TESS follow-up.

The observations were grouped into 12 observing runs  lasting from
0.5 to 3 nights each. The time allocated in 2021 through NOIRLab and Chilean
TACs amounted to 10 nights, two more nights were contributed by the SOAR
partners for TESS follow-up, and some engineering time (morning hours of 8 nights equivalent to 3 full nights) was also used for speckle observations.

%-------------------------------------------------------------
\subsection{Instrument and Observing Procedure}
\label{sec:inst}

The   observations  reported   here  were   obtained  with   the  {\it
  high-resolution camera} (HRCam)  --- a fast imager  designed to work
at  the  4.1  m  SOAR  telescope  \citep{HRCAM}.  The  instrument  and
observing  procedure are  described in  the previous  papers of  these
series  \citep[e.g.][]{SAM19}, so  only the  basic facts  are reminded
here.  HRCam receives  light through  the SOAR  Adaptive Module  (SAM)
which  provides correction  of  the atmospheric  dispersion.  We  used
mostly the near-infrared $I$  filter (824/170\,nm) and the Str\"omgren
$y$  filter  (543/22\,nm), with  four  observations  made in  the  $V$
filter;  the transmission  curves of  HRCam filters  are given  in the
\href{https://noirlab.edu/science/sites/default/files/media/archives/documents/scidoc1740.pdf}{instrument
  manual.}   In  the  standard  observing  mode,  two  series  of  400
200$\times$200  pixel images  (image cubes)  are recorded.   The pixel
scale is 0\farcs01575, so the field of view is 3\farcs15; the exposure
time is normally  24\,ms. For survey programs such  as TESS follow-up,
we use  the $I$ filter and  a 2$\times$2 binning, doubling  the field.
Pairs  wider than  $\sim$1\farcs4 are  observed with  a 400$\times$400
pixel field, and the widest pairs are sometimes recorded with the full
field of 1024 pixels (16\arcsec) and a 2$\times$2 binning.

The  speckle power  spectra are  calculated and  displayed immediately
after       acquisition    for    quick    evaluation    of    the
results. Observations  of close pairs are  accompanied by observations
of single (reference)  stars to account for  such instrumental effects
as telescope vibration  or aberrations.  Bright stars  can be resolved
and measured below the formal diffraction  limit by fitting a model to
the  power  spectrum and  using  the  reference.  The  resolution  and
contrast limits  of HRCam are  further discussed  in TMH10 and  in the
previous papers  of this  series. The standard  magnitude limit  is $I
\approx 12$ mag under typical seeing; pairs as faint as $I \approx 16$
mag were measured under exceptionally good seeing, albeit with reduced
accuracy and resolution.

Custom software helps to optimize observations by selecting targets,
pointing the telescope, and  logging. Typically, about 300 targets are
covered on  a clear night. The  observing programs are  executed in an
optimized  way, depending on  the target visibility,  atmospheric conditions,
and priorities, while minimizing  the telescope slews. Reference stars
and  calibrator binaries are  observed alongside  the main  targets as
needed; their observations are published here as well.

%-------------------------------------------------------------
\subsection{Data Processing}
\label{sec:dat}

The data processing  is described in TMH10 and  \citet{HRCAM}.  We use
the standard speckle interferometry technique based on the calculation
of the power spectrum and the speckle autocorrelation function (ACF).
Companions  are detected  as  secondary  peaks in  the  ACF and/or  as
fringes in  the power spectrum.   Parameters of the binary  and triple
stars  (separation  $\rho$,  position angle  $\theta$,  and  magnitude
difference  $\Delta  m$)  are  determined by  modeling  (fitting)  the
observed  power  spectrum.   The  true  quadrant  is  found  from  the
shift-and-add  (SAA) images  whenever  possible  because the  standard
speckle interferometry determines position angles modulo 180\degr. The
resolution and detection limits are estimated for each observation.

%-------------------------------------------------------------
\subsection{Calibration of Scale and Orientation}
\label{sec:cal}

Since 2014, the  pixel scale and angular offset of  HRCam are inferred
from  observations  of  several  relatively  wide  (from  0\farcs5  to
3\arcsec) calibration binaries, called  {\em calibrators} for brevity.
Their motion is  accurately modeled based on  previous observations at
SOAR  \citep{TMH15}. Before  2014, the  HRCam calibration  was derived
from special experiments described in TMH10 and from the comparison of
HRCam with  the wide-field imager,  SAMI, attached to another  port of
SAM.  The scale and orientation  of both imagers were mutually related
by mapping the motion of a  point source located at the SAM focal plane
and mounted  on the  translation stages. The  orientation of  SAMI was
determined from the astrometric solutions of sky images.

We revised here the list of calibrators and their models using all SOAR
observations from  2007.7 to 2021.75.  After removing faint  stars and
binaries  with known  or  suspected subsystems,  the  list counts  103
calibrators. The motion  of 88 calibrators is modeled  by their visual
orbits,  specially   adjusted  to   accurately  represent   the  HRCam
measurements  (many  of  those  orbits  are  of  low  grade  owing  to
the insufficient coverage); the  remaining 15 pairs are  modeled by linear
functions of the separation $\rho$ and position angle $\theta$ vs. time.

\begin{figure}
\epsscale{1.1}
\plotone{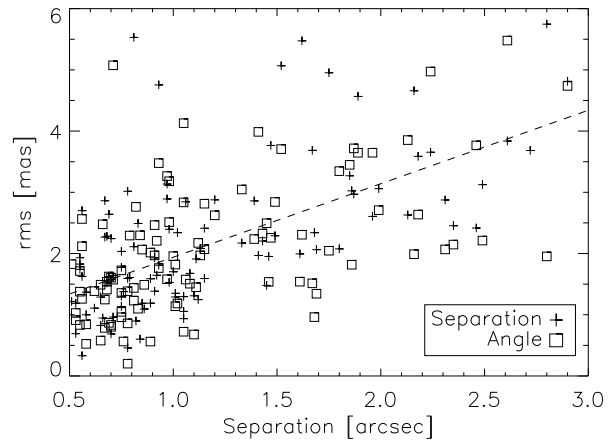}
%\plotone{rms-sep.ps}
\caption{Dependence  of  the  rms  residuals  of  calibrators  on  the
  separation. Plus signs  and squares correspond to  radial $\Delta \rho$
  and tangential $\rho \Delta \theta$  residuals, respectively, and the dashed line is a
  linear fit to both.
\label{fig:rms} }
\end{figure}

Using  the new  models of  the calibrators,  small corrections  of the
angular offset  and scale for  each observing run were  recomputed and
applied. Figure~\ref{fig:rms} plots the rms residuals $\sigma$ between
observations  and  models  in  the radial  and  tangential  directions
vs. separation $\rho$.  The residuals  in both directions are similar,
and they can be jointly approximated by the linear formula
\begin{equation}
\sigma \approx 0.81 + 1.15 \rho  \;\;\; {\rm mas}.
\label{eq:sigma-sep}
\end{equation}
Thus, a 1\arcsec ~binary is measured by HRCam with a typical accuracy
of 2\,mas. A linear increase  of measurement errors with separation is
expected from the physics of  light propagation through the atmosphere
(differential  tilts). Equation 18  from \citet{Kenyon2006}  with
parameters  $C_1  =  500$\,arcsec\,rad$^{-1}$  \,m$^{2/3}$\,s$^{1/2}$,
typical for  Cerro Pach\'on,  binary separation of  1\arcsec, exposure
time of 8  s, and baseline of 2\,m (half  telescope diameter) predicts
the  atmospheric error  of  0.5\,mas, less  than  2\,mas according  to
(\ref{eq:sigma-sep}) and roughly matching the lower envelope of points
in  Figure~\ref{fig:rms}.  Note,  however, that  the calibrator's  model
errors and  the residual  calibration errors  of individual  runs also
contribute to $\sigma$ and  these contributions increase linearly with
$\rho$.  Calibrators with a large  $\Delta m$ have larger residuals to
models.

\begin{figure}[ht]
\epsscale{1.1}
\plotone{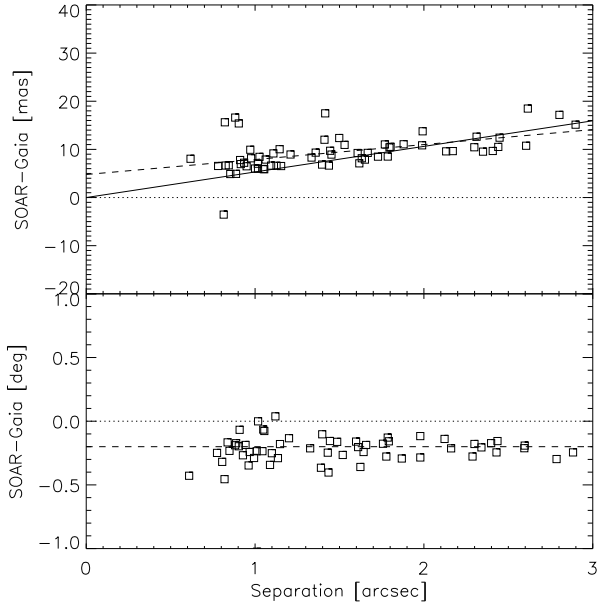}
%\plotone{gaiatestall.ps}
\caption{Dependence of the difference between SOAR and Gaia positions in
  separation (top) and angle (bottom) on the separation. In the top
  plot, the dashed line is a  linear fit and the solid line is a scale
  factor. 
\label{fig:gaia} 
}
\end{figure}

The  set  of  calibrators  assures  a  good  consistency  between  the
observing runs,  but might contain  small global offsets in  angle and
scale, being based entirely on the HRCam data.  An external comparison
is  needed to  control the  systematics. Here  we compare  the updated
calibrator models with  the relative positions of  these pairs derived
from the Gaia Early Data  Release 3 (EDR3) astrometry \citep{Gaia} and
referring to the nominal epoch 2016.0. For some closer pairs, the Gaia
positions strongly  deviate from  the 2016.0 positions  inferred from
the models,  suggesting errors in  the Gaia reductions.  In  most such
pairs with  bad or suspicious  Gaia positons, there are  no parallaxes
and proper motions (PMs) for one or both components.  The remaining 58 pairs
demonstrate  good  consistency  between  Gaia  positions  and  models,
allowing us to look for systematic differences.

The comparison between  the good  Gaia EDR3 calibrators  and their  models is
shown in Figure~\ref{fig:gaia}.   A simple linear fit  to the separation
differences gives
\begin{equation}
\rho_{\rm SOAR} - \rho_{\rm EDR3} \approx 4.8 + 3.11\rho \;\;\; {\rm mas},
\label{eq:rhofit}
\end{equation}
as shown  by the  dashed line in  the top plot.   The existence  of an
offset at  zero separation is not  expected; it could result  from the
errors of the fit.  If real, this  is probably an artifact of the Gaia
measurement algorithm related to blended sources (a similar offset was
found in  the Gaia DR2 positions).   If only the 26  calibrators wider
that 1\farcs5 are selected, the  mean scale factor is $\rho_{\rm SOAR}
/  \rho_{\rm EDR3}  = 1.0053$,  and its  rms scatter  is 0.0009.   The
formal error of  the mean scale factor is thus  $10^{-4}$.  The bottom
panel  of Figure~\ref{fig:gaia}  clearly shows  an offset  between the
position  angles  of   Gaia  EDR3  and  SOAR.   The   mean  offset  is
$\theta_{\rm SOAR} - \theta_{\rm EDR3} = -0\fdg21$, the rms scatter is
0\fdg095, and the formal error of the mean is 0\fdg02.

In  the  light  of  the  comparison  of  calibrators  with  Gaia,  the
separations measured  by HRCam  so far  must be  divided by  the scale
factor  of 1.0053  (decrease by  half a  percent),  and all  position
angles  must  be  increased  by  0\fdg2.   These  corrections  have  a
negligibly small effect  on the orbits of close  binaries derived from
our data, but are  relevant when a small wobble in  the motion of wide
pairs is  studied and when  the HRCam data  are used jointly  with the
Gaia relative positions.  These corrections  are applied to the models
of the calibrators and to the measurements published here. 

Our strategy of observing wide  pairs with slow motion for retroactive
calibration, adopted from  the beginning of the  SOAR speckle program,
has proven  to be correct. Although  the orbits of most  such binaries
are not  of high quality,  they can accurately represent  the observed
motion and serve for comparison with Gaia. The alternative strategy of
using good-quality  orbits of  fast and close  pairs, adopted  by some
other   speckle   programs   \citep[e.g.][]{Horch2021},   delivers   a
substantially inferior  calibration precision.  For example, a  100 mas
pair measured with  a 1 mas accuracy cannot  constrain the calibration
to better than 1\%, even if its orbit were known exactly.

%---------------------------------------------------------
\section{Results}
\label{sec:res}

\subsection{Data Tables}

The  results  (measures of  resolved  pairs  and non-resolutions)  are
presented in  exactly the same format  as in \citet{SAM20}.  The  long tables
are published in machine-readable format; here we describe their content.

\begin{deluxetable}{ l l  l l }
\tabletypesize{\scriptsize}
\tablewidth{0pt}
\tablecaption{Measurements of Double Stars at SOAR 
\label{tab:measures}}
\tablehead{
\colhead{Col.} &
\colhead{Label} &
\colhead{Format} &
\colhead{Description, units} 
}
\startdata
1 & WDS    & A10 & WDS code (J2000)  \\
2 & Discov.  & A16 & Discoverer code  \\
3 & Other  & A12 & Alternative name \\
4 & R.A.     & F8.4 & R.A. J2000 (deg) \\
5 & Decl.    & F8.4 & Declination J2000 (deg) \\
6 & Epoch  & F9.4 & Julian year  (yr) \\
7 & Filt.  & A2 & Filter \\
8 & $N$    & I2 & Number of averaged cubes \\
9 & $\theta$ & F8.1 & Position angle (deg) \\
10 & $\rho \sigma_\theta$ & F5.1 & Tangential error (mas) \\
11 & $\rho$ & F8.4 & Separation (arcsec) \\
12 &  $\sigma_\rho$ & F5.1 & Radial error (mas) \\
13 &  $\Delta m$ & F7.1 & Magnitude difference (mag) \\
14 & Flag & A1 & Flag of magnitude difference\tablenotemark{a} \\
15 & (O$-$C)$_\theta$ & F8.1 & Residual in angle (deg) \\
16 & (O$-$C)$_\rho$ & F8.3 & Residual in separation (arcsec) \\
17  & Ref  & A9   & Orbit reference\tablenotemark{b} 
\enddata
\tablenotetext{a}{Flags: 
q -- the quadrant is determined; 
* -- $\Delta m$ and quadrant from average image; 
: -- noisy data or tentative measures. }
\tablenotetext{b}{References  are provided at
  \url{https://www.astro.gsu.edu/wds/orb6/wdsref.txt} }
\end{deluxetable}

\begin{deluxetable}{ l l  l l }
\tabletypesize{\scriptsize}
\tablewidth{0pt}
\tablecaption{Unresolved Stars 
\label{tab:single}}
\tablehead{
\colhead{Col.} &
\colhead{Label} &
\colhead{Format} &
\colhead{Description, units} 
}
\startdata
1 & WDS    & A10 & WDS code (J2000)  \\
2 & Discov.  & A16 & Discoverer code  \\
3 & Other  & A12 & Alternative name \\
4 & R.A.     & F8.4 & R.A. J2000 (deg) \\
5 & Decl.    & F8.4 & Declination J2000 (deg) \\
6 & Epoch  & F9.4 & Julian year  (yr) \\
7 & Filt.  & A2 & Filter \\
8 & $N$    & I2 & Number of averaged cubes \\
9 & $\rho_{\rm min}$ & F7.3 & Angular resolution (arcsec)  \\
10&  $\Delta m$(0.15) & F7.2 & Max. $\Delta m$ at 0\farcs15 (mag) \\
11 &  $\Delta m$(1) & F7.2 & Max. $\Delta m$ at 1\arcsec (mag) \\
12 & Flag & A1 & : marks noisy data  
\enddata
\end{deluxetable}

\begin{deluxetable*}{l l}    
\tabletypesize{\scriptsize}     
\tablecaption{Notes
\label{tab:notes}          }
\tablewidth{0pt}                                   
\tablehead{                                                                     
\colhead{WDS} & 
\colhead{Comment} 
}
\startdata
02346$-$4210& This is WDSS 0234367$-$421019 (HIP 11989), see 2020ApJS..247...66H \\
02434$-$6643& B is fainter than A in $y$ filter, but brighter in $I$. \\
02415$-$7128& HIP 12548 is a quadruple system, orbits in  2022AJ....163..161T  \\
03260$-$3558& B 1449BC is a triple in a young (40 Myr) cluster, first 41 yr orbit here. \\
03284+2248& The subsystem BAG 2Aa,Ab is not detected in 2018-2021, spurious? \\
03478$-$1854& This is WDSS 0347502$-$185407 (TIC 121088959), see 2020ApJS..247...66H \\
04293$-$3124& SIG 4 is a brown dwarf. Noisy measures, poor accuracy. Orbit $P=55$ yr.  \\
04375+1509& CHR 153 contains a subsystem, orbits in 2021AJ....161..144T \\
04404+1631& CHR 154 is a triple in the Hyades, orbits in 2021AJ....161..144T \\
05303$-$6653& A small $\Delta I=-0.06$ mag an does not match $\Delta G=0.93$ mag,  variable? \\ 
05321$-$0305&  The resolution of V1311Ori Ba,Bb is reported in   2022AJ....163..127T  \\
05351$-$0249& A and B components of SKF2259 (Haro 5-2, a PMS star) were resolved   \\
          & into subsystems by Bo Reipurth (2022, in preparation).  \\
06035+1941& MCA 24 is a B8III triple with a 14-day subsystem. The outer orbit is updated here. \\
06314+0749& Triple system, orbits in 2020AJ....160...69T  \\ 
07498$-$0317& This is WDSS 0347502$-$185407, see 2018MNRAS.480.4884E \\
08240$-$1548& HIP 41171 is a quadruple system, see 2019AJ....158..222T \\
11100$-$6645& The resolution is reported by Powell et al. 2021AJ....162..299P \\
          & This is a TESS object with enigmatic eclipses, apparently by dust. \\
11221$-$2447& This is a PMS quadrule HD 98800, orbits in 2021A\&A...655A..15Z \\
13175+2024& The subsystem in YSC~129 (HIP 64386) belongs to A, not to  B. \\
          &  Preliminary orbits indicate periods of 34 and 5 yr for AB and Aa,Ab, respectively. \\
15234$-$5919& B is red, fainter than A in the $V$ band but brighter in $I$. \\
16243$-$5921& The measured pair (0\farcs99, $\Delta I=4.2$) is found in Gaia EDR3 at the same position. \\
          & The parallax is 0.17\,mas, so it cannot be the 0\farcs3 pair HDS2317Aa,Ab \\
17289$-$3244& AB (2\farcs36) is WDSS 1728556$-$324356, BC is a new pair, see the text. \\
%          & detected by Bonavita et al.,  http://arxiv.org/abs/2103.13706 \\
19250$-$3205& AB is WDSS 1924594$-$320432 (TIC 11247221), see  2020ApJS..247...66H; \\
          & The 0\farcs08  pair Aa,Ab is new.  \\
21145$-$7403& This is WDSS 2114277$-$74024 (HIP 104556), see 2021MNRAS.506.2269E 
\enddata 
\end{deluxetable*}

Table~\ref{tab:measures}  lists  2,623 measures  of 2,123 resolved pairs
and subsystems,  including new discoveries.  The  pairs are identified
by their WDS-style codes based on the J2000 coordinates and discoverer
designations adopted  in the  WDS catalog \citep{WDS},  as well as by
alternative   names  in   column  (3),   mostly  from   the  Hipparcos
catalog.  Equatorial coordinates for  the epoch  J2000 in  degrees are
given  in  columns (4)  and  (5)  to  facilitate matching  with  other
catalogs and databases.  In the case of resolved multiple systems, the
position measurements  and their errors (columns  9--12) and magnitude
differences  (column  13) refer  to  the  individual pairings  between
components, not to their photocenters.   As in the previous papers of
this  series, we  list  the  internal errors  derived  from the  power
spectrum model  and from the difference between  the measures obtained
from two data  cubes.  The real errors are  usually larger, especially
for  difficult pairs  with substantial  $\Delta m$  and/or  with small
separations.    Residuals  from   orbits  and   from  the   models  of
calibrators,  typically between  1  and 5  mas  rms, characterize  the
external errors of the HRcam astrometry (Figure~\ref{fig:rms}).

The  flags in column  (14) indicate the cases where  the true  quadrant is
determined (otherwise the position angle is measured modulo 180\degr),
when the  relative photometry of wide  pairs is derived from  the long-exposure
images (this  reduces the bias  caused by speckle  anisoplanatism), and
when the data are noisy  or the resolutions are tentative (see TMH10).
For binary stars with known  orbits, the residuals to the latest orbit
and its reference are provided in columns (15)--(17). Residuals close
to 180\degr ~mean that the orbit swaps the brighter (A) and fainter
(B) stars. However, in some binaries the secondary is fainter in one
filter and brighter in other (e.g. 15234$-$5919). In these cases, it is
better to keep the historical identification of the components in
agreement with the orbit and to give a negative magnitude difference
$\Delta m$. 

Nonresolutions  are  reported  in Table~\ref{tab:single}.  Its  first
columns  (1)  to   (8)  have  the  same  meaning   and  format  as  in
Table~\ref{tab:measures}.  Column  (9)  gives the  minimum  resolvable
separation when  pairs with $\Delta m  < 1$ mag are  detectable. It is
computed from  the maximum spatial  frequency of the useful  signal in
the power  spectrum and  is normally close  to the  formal diffraction
limit  $\lambda/D$. The following  columns (10)  and (11)  provide the
indicative  dynamic range,  i.e. the  maximum magnitude  difference at
separations  of 0\farcs15  and 1\arcsec,  respectively, at  $5\sigma$
detection level.  The last  column (12) marks  noisy data by  the flag
``:''.
In Table~\ref{tab:notes} we provide short notes on some pairs, mostly
the alternative designations and relevant references. 

%---------------------------------------------------------
\subsection{New Pairs}
\label{sec:new}

\begin{figure}
\epsscale{1.1}
\plotone{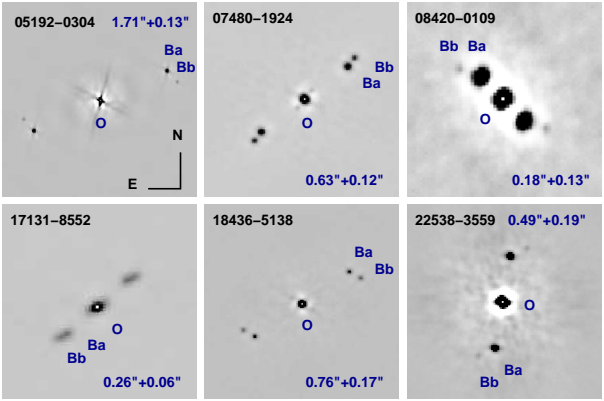}
%\plotone{Triples.eps}
\caption{Fragments  of  speckle ACFs  of  some  newly resolved  triple
  stars.   The  intensity  and  spatial scales  are  chosen  for  best
  representation of each system. North  is up, east left. Blue letters
  mark the  ACF peaks  corresponding to  the component's  location (as
  opposed to other symmetric peaks), O  marks the ACF center. The
  outer and inner separations are indicated.
\label{fig:trip} }
\end{figure}

\startlongtable

\begin{deluxetable}{ l l c  c l  }
\tabletypesize{\scriptsize}
\tablewidth{0pt}
\tablecaption{New Double Stars
\label{tab:binaries}}
\tablehead{
\colhead{WDS} &
\colhead{Name} &
\colhead{$\rho$} &
\colhead{$\Delta m$} &
\colhead{Program\tablenotemark{a}} \\
\colhead{J2000}  &     &    
 \colhead{(arcsec)} &
 \colhead{(mag)} & 
}
\startdata
00272$-$7324 &    TOK 908 & 0.24 & 1.7 & HIP \\
00438$-$3911 &    TOK 909 & 0.66 & 3.6 & HIP \\
01333$-$5142 &    TOK 910 & 0.36 & 2.6 & HIP \\
02035$-$3005 &     RST 2270 Aa,Ab & 0.22 & 3.1 & MSC  \\ % TRI Pow
03230$-$7047 &    HEI 630 BC & 0.12   & 0.8 & MSC  \\ % TRI Pow
03385$+$1336 & YR 10 Ab1,Ab2 & 0.04 & 0.7 & MSC\tablenotemark{b}  \\ % TRI
05053$-$4208 &  BRT 668 Aa,Ab & 0.15 & 2.2 & TESS \\ % TRI
05192$-$0304 &   A 53 Ba,Bb  & 0.37 & 1.9 & MSC\tablenotemark{b} \\ % TRI 2021AJ....161..134H dK=2.77
05427$-$7316 & HEI 667 AC    & 1.35 & 4.4 & NEG \\ %
06005$-$2753 & HDS 819 Ba,Bb & 0.09 & 1.6 & NEG \\ % TRI
06159$-$5126 & RST 179 BC    & 0.07 & 1.0 & MSC \\ % TRI Pow
06223$-$2021 & TDS 3781 BC    & 0.12 & 1.4 & MSC \\ % TRI Pow
07018$-$1118 & HU 112 Ba,Bb  & 0.04 & 0.1 & MSC\tablenotemark{b}  \\ % TRI  wobble
08321$-$2730 & HDS 1220 BC   & 0.04 & 0.2 & NEG \\ % TRI
08420$-$0109 &  A 1750 Ba,Bb & 0.12 & 2.7 & NEG\tablenotemark{b} \\ % TRI MSC
08471$-$4504 & TOK 911     & 0.04 & 1.0 & REF \\
09046$-$4104 & RSS 189 Aa,Ab    & 0.06 & 0.3 & HIP\tablenotemark{b}  \\
09343$-$3223 & RST 2637 Aa,Ab & 0.22 & 3.3 & MSC \\ % TRI Pow
10350$-$5247 & B 1685 BC     & 0.40 & 2.4 & TESS \\ % TRI
11247$-$6139 & BSO 5 Aa,Ab   & 0.25 & 2.8 & MSC \\
12059$+$2628 & HDS 1707 Ba,Bb & 0.07 &  1.5 &  HIP \\ % TRI
12118$-$5306 & HDS 1722 AB    & 0.83 & 5.6  & WDS \\ % Add 2018  
12282$+$2502 & TOK 912      & 0.03 & 0.7 & HIP \\ %
13514$+$2620 & SKF 260 Ba,Bb  & 0.25 & 0.2 &  MSC\tablenotemark{b}  \\ %  2019 Bab not meas). quadr
13592$-$4528 & TOK 913     & 1.20 & 4.6 & REF \\ % Also EDR3. opt?
17131$-$8552 & JNN 320 Ba,Bb  & 0.06 & 0.1 & YMG\tablenotemark{b}  \\ %  TRI
17289$-$3244 & TOK 914 BC & 0.32 & 2.0 & YMG\tablenotemark{b} \\ % Also Bonavita2021
17373$-$4300 & SEE 510 Aa,Ab & 0.24 & 3.7 & WDS \\ % R.Argyle, AB does not exist
17374$-$3544 & HDS 2488 AC     &  2.24 & 5.3 & NEG \\ % opt, Gaia
18153$-$4928 & TOK 915   & 0.07 & 0.0 & YMG \\
18248$-$0621 & TOK 320 Ba,Bb & 0.13 & 1.3 & MSC \\
18285$-$4129 & BRT 1058 Aa,Ab & 0.17 & 0.2 &  YMG\tablenotemark{b} \\ % AB 6'' phys, to MSC
18436$-$5138 & RST 5451 Ba,Bb & 0.18 & 0.3 &  NEG \\ % TRI
18483$-$3710 & JSP 787 Aa,Ab  & 0.59 & 3.3 & MSC \\ %  Pow
18581$-$2953 & SKF 1507 Aa,Ab & 0.07 & 1.6 & YMG \\  %2017A&A...599A..70J Janson, Eliott
18581$-$2953 & SKF 1507 Ba,Bb & 0.37 & 0.0 & YMG\tablenotemark{b}  \\  %2017A&A...599A..70J Janson, Eliott
19250$-$3205 & TOK 916 Aa,Ab & 0.08 & 0.9 & YMG \\ % TRI TOK 916?
20008$-$2411 & TOK 917       & 0.70 & 2.6 & HIP \\  
20446$-$6630 & TOK 918      & 0.08 & 0.1 & HIP \\
20557$-$6609 & HDS 2981 AC    & 1.32 & 6.3 & WDS \\  % ORB in fact
21433$-$4327 & HDS 3095 Ba,Bb  & 0.11 & 3.1 & HIP \\ % 2014.7 UR, CTE pbm
21510$+$2911 & A 889 Aa,Ab     & 0.03 & 0.4 & MSC \\ % wobble?
22039$-$2451 & SEE 465 Aa,Ab   & 0.65 & 1.7 & MSC\tablenotemark{b} \\ % GKM
22039$-$2451 & SEE 465 Ba,Bb   & 0.15 & 1.9 & MSC\tablenotemark{b} \\ % GKM
22146$-$2142 & TOK 919        & 0.09 & 1.1 & MSC \\ % sb4
22538$-$3559 & HDS 3255 Ba,Bb  & 0.19 & 2.1 & NEG \\ % TRI
23214$-$1340 & HDS 3326 Aa,Ab  & 0.05 & 0.0 & NEG \\ % TRI
23224$-$5857 & TOK 920     & 0.07 & 0.1 & HIP \\
23293$-$5135 & TOK 921     & 0.25 & 2.7 & HIP \\
23465$-$4135 & TOK 922     & 0.74 & 3.4 & HIP 
\enddata 
\tablenotetext{a}{
HIP -- Hipparcos suspected binary; 
MSC -- multiple system; 
REF -- reference star; 
YMG --  young moving groups;  
NEG -- neglected pair; 
TESS -- TESS followup.
}
\tablenotetext{b}{See comments in the text.}
\end{deluxetable}

%02346$-$4210 &    HIP 11989  & 2.73   & 3.8 &  MSC \\ % in EDR3 Pow 2020ApJS..247...66H
%07480$-$1924 & B  1077 Ba,Bb & 0.12 & 0.7 & NEG\tablenotemark{b}  \\ % TRI not in WDS, in SOAR2019
%17289$-$3244 & HD 317617 AB & 2.36 & 2.3 & YMG\tablenotemark{b} \\ % TRI Also Bonavita2021, EDR3
%18248$-$0621 & HIP 90246 Aa,Ab & 0.31 & 2.9 & MSC\tablenotemark{b}  \\ % Quadr
%19250$-$3205 & TIC 11247221 AB  & 2.00 & 0.8 & YMG \\ % TRI
%19336$+$0346 & HIP 96196       & 0.24 & 3.2 & REF \\ % Not in WDS,res.2018 TOK 903
%21145$-$7403 & HIP 104556      & 1.10 & 2.4 & HIP \\ % Gaia as well, WDSS

Table~\ref{tab:binaries} highlights  the 50 pairs resolved  in 2021 or
resolved earlier  but not  yet published.   All measurements  of these
pairs are found in Table~\ref{tab:measures}.  The pairs are identified
by the  WDS-style codes and the  discovery codes or other  names.  The
following  columns  contain  the   separation  $\rho$,  the  magnitude
difference  $\Delta m$,  and the  observing program.   Two pairs  were
found  independently by  others, but  we keep  them in  the table  for
consistency  because they  are  not yet  featured in  the  WDS or  its
supplement, WDSS. The majority of  newly discovered binaries belong to
hierarchical  systems with  three or  more components.  Typically, new
close subsystems  in known visual  binaries were discovered  by HRCam.
Comments on  some multiple  systems are given  below, and  the speckle
ACFs of selected triples are shown in Figure~\ref{fig:trip}.

%02346$-$4210 (HIP 11989). The 2\farcs73 pair is also present in Gaia EDR3.

03385+1336 (HIP  16991, F8) is a  quadruple system at 95  pc distance.
The outer  pair A,B  has a separation  of 77\arcsec,  the intermediate
pair Aa,Ab (YR 10) is at 0\farcs27, and the newly discovered subsystem
Ab1,Ab2 is at  0\farcs04. The estimated period of  Ab1,Ab2 is $\sim$10
yr, and it should leave an imprint on the observed motion of Aa,Ab.

05192$-$0304  (HIP 28419,  K3V).  The subsystem  Ba,Bb  in a  1\farcs5
binary A~53  has been discovered by  \citet{Hirsch2021} using adaptive
optics. They  measured it at  243\fdg4 and 0\farcs520 in  2015.410 and
determined the  magnitude difference of  $\Delta K_{\rm Ba,Bb}  = 1.2$
mag; another observation was made in 2017.931. Our measurement in 2021
shows  that Ba,Bb  has  retrograde  motion and  is  closing down;  its
estimated period  is 25  yr. The  outer pair A,B  had a  separation of
4\farcs9 in 1900, and now it closed down to 1\farcs5. This low-mass triple
system is located at 15.7\,pc from the Sun.

07018$-$1118 (HIP 33868, B2V). The main  star A is an eclipsing binary
GU~CMa  with   a  close  tertiary  companion   discovered  by  eclipse
timing. The  resolution of  Ba,Bb at 0\farcs04  (AB is  at 0\farcs65)
makes this  a quintuple hierarchical  system. The estimated  period of
Ba,Bb is 25 yr. More observations are needed to confirm the close pair
Ba,Bb  and to  follow  its  motion. It  cannot  be  excluded that  the
eclipsing pair belongs to star B, in which case its tertiary component
is identical to Bb and the system is quadruple rather than quintuple.

08420$-$0109 (HD 74113, A6V) is a  strange case. The orbit of the main
pair A,B (A~1750) with a period of  $P=250$ yr and a semimajor axis of
$a  = 0\farcs205$  is  determined  here. The  new  faint star Bb with  a
separation of  0\farcs12 from Ba would  make the system  dynamically unstable,
unless its true separation is  larger than  $\sim$0\farcs6. In such case,
Bb  is  an  outer  companion,  and  Ba,Bb  looks  close  only  in
projection.  The detection  of Ba,Bb  is securely  confirmed in  three
observing runs. This subsystem is  also apparent in our 2018 observation,
although it  was overlooked at  the time because  the ACF was  distorted by
telescope vibration. Gaia  does not give the parallax. The  sky is not
crowded, making  a random superposition  of unrelated stars at  such close
separation  extremely unlikely.  A similar 0\farcs48  pair ADS~6941
(BD$-$00\degr  2045), located  at 137\arcsec  ~from  AB, also lacks  Gaia
parallax and PM, and its relation to this triple is presently unknown.

09046$-$4104 (HIP  44550, G1V)  is a  three-tier hierarchical  system. The
inner pair is an eclipsing binary V405~Vel with a 10 day period. It is
orbited  by star  Ab at  0\farcs06 discovered  here with  an estimated
period of 20  yr. This subsystem explains the  large astrometric noise
in Gaia. The outermost component B, at 9\farcs5 separation, has common
PM and parallax.

13514+2620  (HIP 6723,  K6V)  is a  low-mass 2+2  quadruple  at 46  pc
distance.  Aa,Ab  is  a  known  0\farcs3 pair  YSC~50,  and  Ba,Bb  at
0\farcs25 is discovered here. Both  inner pairs have estimated periods
of $\sim$50 yr and large mass  ratios.  The outer pair A,B at 3\farcs3
has  a  period of  the  order  of 1  kyr.  Gaia  did not  measure  the
parallaxes of A and B because they are close binaries.

17131$-$8552 (GSC  09526$-$00895, M0V)  is an interesting  triple system
where the estimated periods of the outer 0\farcs26 and inner 0\farcs06
pairs are 25  and 3 yr, respectively. All three  stars have comparable
magnitudes. Gaia gives no parallax for this star.
  
17289-3244 (HD  317617, TYC 7379-279-1,  K3V). This triple  system has
been resolved in 2015.41 by \citet{Bonavita2021} in a survey of nearby
young stars. They measured BC  at 318\fdg2, 0\farcs416 with $\Delta
H_{\rm BC} = 1.9$ mag.

17373$-$4300 (HIP 86228, $\theta$~Sco,  F1III) was observed on request
by R.~Argyle because it was suspected to be a close pair in Hipparcos.
The newly resolved high-contrast 0\farcs24  pair is different from the
known 6\farcs2 pair  SEE 510. The latter in fact  is spurious, because
it was  not seen in  the wide-field images  taken with HRCam,  and the
secondary is  not found  in Gaia.  The  Hipparcos measurement  of this
pair at 6\farcs5 separation is  spurious, probably  caused by the
wrongly interpreted signal from the close pair measured here.

18248$-$0621 (HIP 90246, K7V) is another nearby (42\,pc) 2+2 quadruple
system   discovered  here.    The   outer  binary   is  at   55\arcsec
~separation. Its  components A and  B are  revealed to be  close pairs
with separations of  0\farcs31 and 0\farcs13 and  estimated periods of
40 yr and 15 yr,  respectively. The Gaia astrometry of A  and B has large
noise owing to their binary nature, but the Gaia parallaxes and PMs of
A and B confirm that they form a physical system.

18285$-$4129 (TYC 7909-2501-1) is a  physical triple system. The WDS
pair BRT~1058 (5\farcs2, 178\degr, $\Delta m = 2$ mag) is not found in
Gaia, but  a much  fainter ($G  = 15.48$ mag)  Gaia star  at 6\farcs1,
205\fdg0 from A  is a physical companion with common  PM and parallax.
Star  A  is  resolved  here   into  a  0\farcs17  pair,  although  its
designation as BRT~1058Aa,Ab appears misleading.

18581$-$2953 (TYC  6872-1011-1, M0V)  is another nearby  2+2 quadruple
discovered here  by resolving both  components of the  28\arcsec ~pair
into 0\farcs07  and 0\farcs37  subsystems. The  closer one,  Aa,Ab, is
expected to have a short period  of $\sim$10 yr. The system belongs to
a   young   moving  group.   Star   A   was   observed  at   Keck   by
\citet{Ruane2019},  but   this  close  pair  was   hidden  behind  the
coronagraphic mask of 0\farcs25 radius, preventing its detection.

22039$-$2451 (HIP 108923, G6V) contains five components in a hierarchical
configuration and  is at 57\,pc  distance from the Sun.  The outermost
component C  is at  122\arcsec ~from  the 3\farcs4  pair AB  (SEE 465
AB).  We resolved  here both  components of  SEE~465 into  close pairs
Aa,Ab (0\farcs66) and Ba,Bb (0\farcs15)  with estimated periods of 130 yr
and 20  yr, respectively, and substantial  magnitude differences. Gaia
measured matching  parallaxes of A, B,  and C, while their  PMs differ
owing to  the motion in  the inner  subsystems. Although Gaia  did not
resolve  Aa,Ab  explicitly, it  is  marked  as having  double  transits,
indicating the detection of Ab.

Some wide  pairs resolved here are  also found in the  Gaia EDR3. When
parallaxes and PMs of both components  are given, the wide pair can be
classified as  physical (e.g.  13592$-$4528)  or optical
(chance projection), as  17374$-$3544AC.

%-------------------------------------------------------------
\subsection{New and Updated Orbits}
\label{sec:orbits}

Orbits of visual  binary stars is a traditional  battleground of human
ingenuity     against     inaccurate     data     and     insufficient
coverage. Availability of computing  power added to various historical
methods developed during two  centuries such approaches as statistical
sampling \citep[e.g.][]{Mendez2022}  and a brute-force  exploration of
the   multidimensional  parameter   space  \citep{Blunt2020}.    Space
astrometry  and radial  velocities (RVs)  are also  used nowadays  for
orbit calculation  \citep{Brandt2021a}.  This field is  rejuvenated by
recent imaging of substellar and  planetary companions and the need to
infer their orbits from the short observed arcs. The situation will be
further exacerbated  by millions  of binaries  expected in  the future
data  releases  of  Gaia,   e.g.  one  million   binaries
    published  by \citet{ElBadry2021},  and lacking  coverage  of  their
orbits. 

Positional measurements provided by the SOAR speckle program contain a
rich material for calculation of  new visual orbits and improvement of
the known ones. Some orbits  resulting from this program are published
in \citep{Mendez2022,Gomez2022, Tok2021a} and  in the circulars of the
IAU Commission G1.  The quality of these orbits  ranges from tentative
and  preliminary (grades  5 and  4) to  good, reliable,  and excellent
(grades 3  to 1). Orbits  are graded using the code and methodology
described in ORB6 \citep{VB6}. 

Table~\ref{tab:vborb} lists elements  of  174 new and
corrected  orbits determined  from the  observations made  at SOAR  in
2021.   The complementary  Table~\ref{tab:vborb2} contains  additional
 108 preliminary orbits where no reliable estimates of
the element's  errors could be  derived. Most preliminary  orbits have
grades 4 and 5.  The constraints  on these orbits are insufficient for
meaningful estimates of the elements' errors, and some of these orbits
will be  improved or even  dramatically revised  in the future  as new
data  are accumulated.   Calculation and  improvement of  orbits is  a
never ending process where orbital solutions progressively become more
reliable  and accurate,  while  new low-grade  preliminary orbits  are
being added to the catalog.  Note however that some grade 4 orbits are
placed   in  Table~\ref{tab:vborb}   because  the   errors  are   well
defined.  Conversely,  some  orbits  of official  grade  3  are  still
considered  preliminary  and   placed  in  Table~\ref{tab:vborb2}.   A
one-dimensional grading system  cannot grasp all aspects  of the orbit
quality; it is useful for a  first-order evaluation. The errors of the
elements  and the  plots  provided in  the ORB6  catalog  give a  more
comprehensive picture.

Most orbits were fitted by the weighted least-squares method using the
IDL program  {\tt ORBIT} \citep{orbit}.  The  weights are proportional
to  $\sigma^{-2}$, and  the measurement  errors $\sigma$  are assigned
depending on  the data source: 2  to 5 mas for  speckle interferometry
with telescopes of  4 m class and  Gaia, 10 mas for  Hipparcos, 50 mas
and  larger  for visual  micrometer  measures.   Some outlying  visual
measures  are  ignored.   We  also use  the  radial  velocities  where
available. Some  orbits were  fitted using  the grid  search algorithm
\citep{Hartkopf1989}.   Different systems  of weights  are adopted  by
other orbit computers; usually more  weight is given to the historical
micrometer data at the expense of  a worse fit to the accurate speckle
positions.

The orbital elements and their errors (determined by the least-squares
fitting)   are  given   in  Table~\ref{tab:vborb}   in  the   standard
notation. The last  columns contain the grade and  the reference codes
to   previously   published   orbits,  when   available.    The   full
bibliographic     references    are     provided    in     the    ORB6
\href{http://www.astro.gsu.edu/wds/orb6/wdsref.txt}{online
  catalog}. 

\startlongtable

\begin{deluxetable*}{r l cccc ccc cc}    
\tabletypesize{\scriptsize}     
\tablecaption{Visual Orbits with Element's  Errors [Fragment]
\label{tab:vborb}          }
\tablewidth{0pt}                                   
\tablehead{                                                                     
\colhead{WDS} & 
\colhead{Discov.} & 
\colhead{$P$} & 
\colhead{$T$} & 
\colhead{$e$} & 
\colhead{$a$} & 
\colhead{$\Omega$ } & 
\colhead{$\omega$ } & 
\colhead{$i$ } & 
\colhead{Grade }  &
\colhead{Ref.\tablenotemark{a}} \\
 \colhead{{\it HIP}} &
& 
\colhead{(yr)} &
\colhead{(yr)} & &
\colhead{(arcsec)} & 
\colhead{(deg)} & 
\colhead{(deg)} & 
\colhead{(deg)} &  & 
%\colhead{} &
%\colhead{} & 
}
\startdata
00219$-$2300 & RST5493 BC & 18.441 & 2019.527 & 0.831 & 0.1338 & 91.3 & 213.6 & 92.1 & 3 & Tok2021c \\
             &     & $\pm$0.555 & $\pm$0.447 & $\pm$0.034 & $\pm$0.0204 & $\pm$1.5 & $\pm$16.0 & $\pm$1.7&     &  \\
00277$-$1625 & YR 1 Aa,Ab & 6.586 & 2014.311 & 0.762 & 0.0675 & 167.2 & -0.9 & 18.5 & 2 & Tok2020e \\
             &     & $\pm$0.027 & $\pm$0.035 & $\pm$0.015 & $\pm$0.0012 & $\pm$29.9 & $\pm$31.5 & $\pm$9.0&     &  \\
00324$+$0657 & MCA 1 Aa,Ab & 27.506 & 1989.001 & 0.810 & 0.1593 & 105.8 & 14.3 & 110.8 & 2 & Jte2018 \\
             &     & $\pm$0.051 & $\pm$0.105 & $\pm$0.025 & $\pm$0.0022 & $\pm$0.8 & $\pm$2.3 & $\pm$2.3&     &  \\
01077$-$1557 & HDS 148 & 15.493 & 2019.120 & 0.980 & 0.0738 & 11.0 & 213.0 & 66.8 & 3 & Tok2021c \\
             &     & $\pm$0.755 & $\pm$0.207 & fixed & $\pm$0.0152 & $\pm$7.1 & $\pm$20.3 & $\pm$11.1&     &  
\enddata 
\tablenotetext{a}{References to previous orbits are provided in 
\href{http://www.astro.gsu.edu/wds/orb6/wdsref.txt}{wdsref.txt}.
Additional references are:
%Bonavita2021 -- \citet{Bonavita2021};
Circ204 -- Circular of IAU Commission G1, No. 204, 2021;
Gomez2022 -- \citet{Gomez2022};
Horch2021 -- \citet{Horch2021};
%Mann2019 -- \citet{Mann2019};
%Mdz2022 -- \citet{Mendez2022};
%Vri2022-- \citet{Vrijmoet2022}.
 }
\end{deluxetable*}

\startlongtable
\begin{deluxetable*}{l l cccc ccc cc}    
\tabletypesize{\scriptsize}     
\tablecaption{Preliminary Visual Orbits  [Fragment]
\label{tab:vborb2}          }
\tablewidth{0pt}                                   
\tablehead{                                                                     
\colhead{WDS} & 
\colhead{Discov.} & 
\colhead{$P$} & 
\colhead{$T$} & 
\colhead{$e$} & 
\colhead{$a$} & 
\colhead{$\Omega$ } & 
\colhead{$\omega$ } & 
\colhead{$i$ } & 
\colhead{Grade }  &
\colhead{Ref.\tablenotemark{a}} \\
 & & 
\colhead{(yr)} &
\colhead{(yr)} & &
\colhead{(arcsec)} & 
\colhead{(deg)} & 
\colhead{(deg)} & 
\colhead{(deg)} &  & 
%\colhead{} &
%\colhead{} & 
}
\startdata
00258$+$1025 & HDS 57 & 69.179 & 2020.847 & 0.471 & 0.1413 & 43.6 & 59.2 & 137.0 & 4 & Tok2021c \\
00304$-$6236 & JNN 296 Aa,Ab & 60.644 & 2016.792 & 0.664 & 0.3132 & 82.1 & 24.6 & 130.1 & 4 & first \\
00348$-$5853 & I 439 & 350.000 & 2020.615 & 0.960 & 0.7021 & 119.4 & 146.2 & 48.9 & 4 & Tok2021c \\
01114$+$1526 & BEU 2 & 62.568 & 1993.862 & 0.000 & 0.4676 & 120.5 & 0.0 & 51.4 & 4 & Tok2021c \\
01117$+$0835 & HDS 158 & 47.008 & 2023.553 & 0.853 & 0.1985 & 132.0 & 172.6 & 63.8 & 4 & first 
\enddata 
\tablenotetext{a}{See Notes to Table~\ref{tab:vborb} }
\end{deluxetable*}

\begin{figure}
\epsscale{1.1}
\plotone{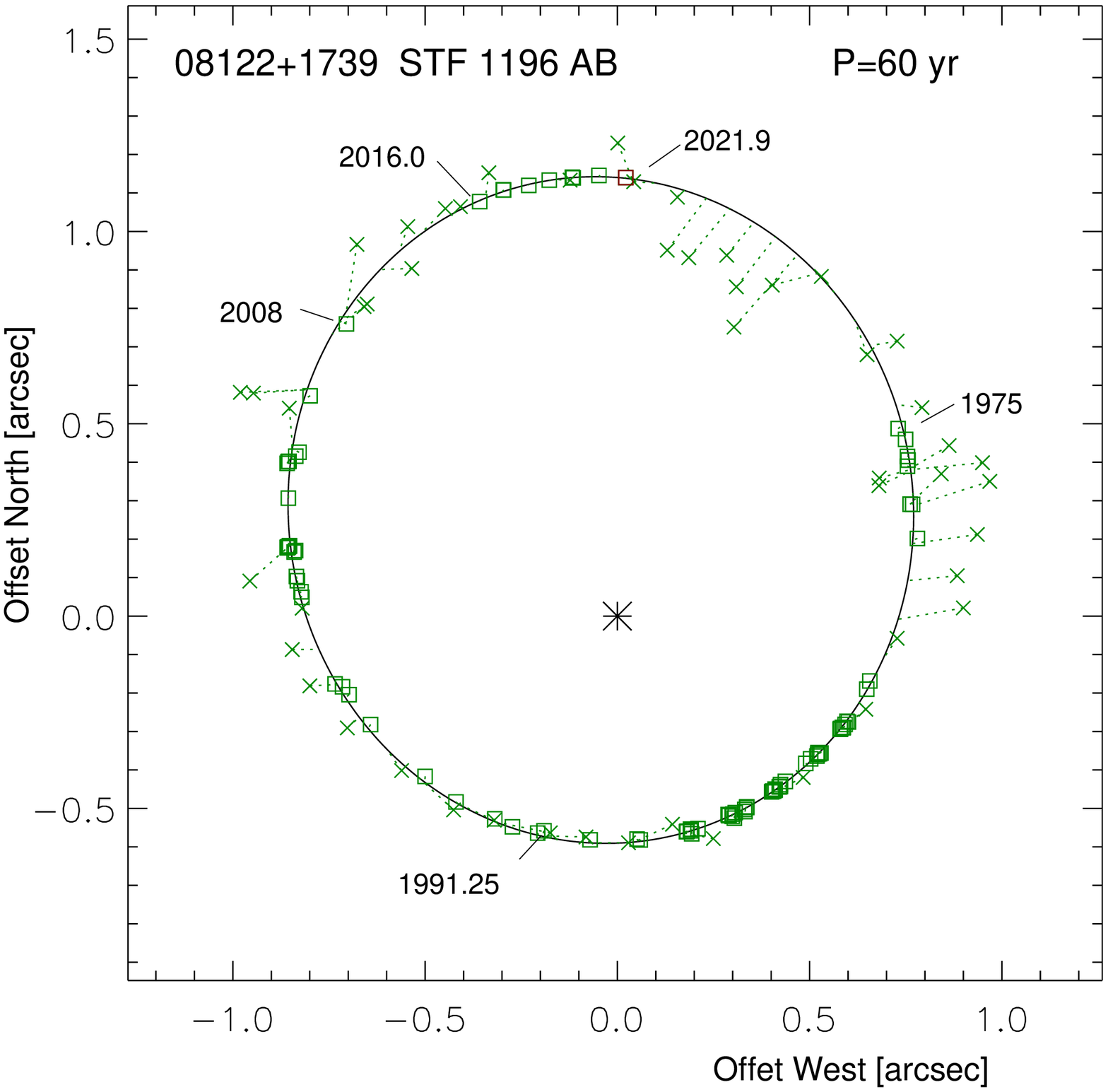}
\plotone{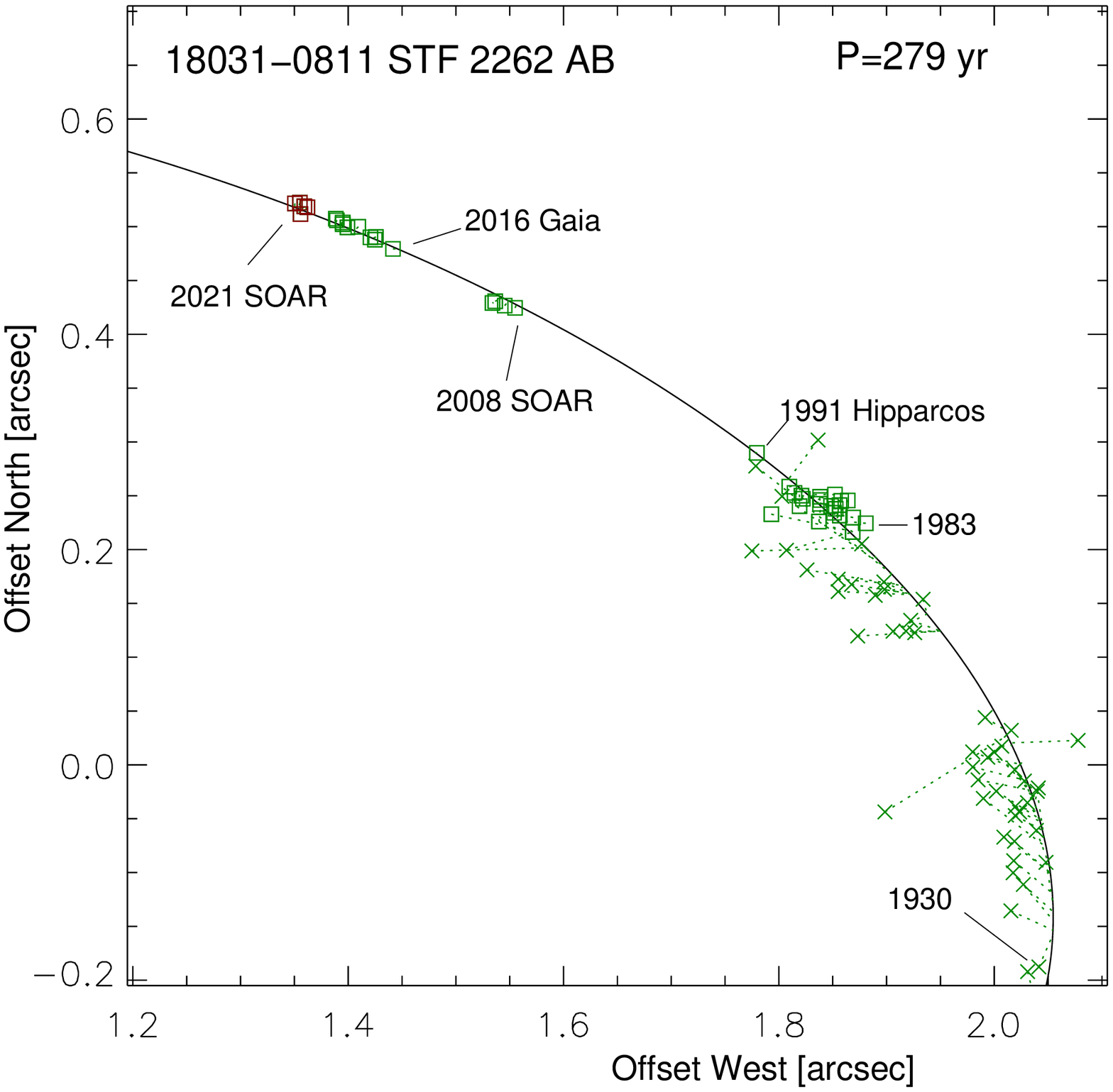}
%\plotone{STF1196.eps}
%\plotone{STF2262.eps}
\caption{Updated orbits of two calibrators, STF~1196 AB (top) and
  STF~2262 AB (bottom). Accurate speckle measurements are plotted as
  squares (red color marks the 2021 observations published here), the
  less accurate micrometric observations as crosses. Dotted lines
  connect the positions to the locations on the orbit (ellipse). The scale
  is in arcseconds, and the main component is at coordinate origin. 
\label{fig:calorb} }
\end{figure}

Figure~\ref{fig:calorb}  shows   updated  orbits  of   two  calibrator
binaries.  Although  the orbit of  STF~1196 ($\zeta$ Cnc AB)  has been
determined   many  times   in   the  past,   its   latest  update   by
\citet{Izm2019} shows appreciable systematic residuals to the accurate
SOAR and Gaia  positions.  This pair has been measured  more than 1000
times since  its discovery by  W.~Struve in 1825, mostly  using visual
micrometers.   However,  fitting  the  orbit to  all  available  data,
suitably weighted, does not lead to a good result because the numerous
but intrinsically inaccurate visual  measurements pull the solution in
the wrong  direction.  Here  we use  only the  micrometer measurements
made by  W. and  O. Struve in  the 19th century  and ignore  all other
measurements except  speckle interferometry  with telescopes of  1.8 m
aperture or larger.  This pair has been frequently observed by speckle
interferometry since  1975 as a  calibrator, and the speckle  data now
cover most  of its  60 yr  orbit.  Our new  accurate orbit  makes this
binary a first class calibrator.  It  is bright, has a small magnitude
difference, and  is accessible  from both  hemispheres.  Incidentally,
the  pair belongs  to the  nearby  (25 pc)  quintuple multiple  system
$\zeta$ Cnc,  and the orbit  of its subsystem  Ca,Cb with a  period of
17.3 yr is  also updated here.  With an accurate  parallax from future
Gaia data releases, both orbits will yield accurate masses.

The bottom panel of Figure~\ref{fig:calorb} shows a more typical
calibrator where accurate measurements cover only a small arc of the
long-period orbit. The elements fitted with appropriate weights
accurately represent this arc. The 18 calibrated SOAR measurements
have  rms residuals of 3.6 mas and agree very well with the Gaia EDR3
position.

\begin{figure*}
\epsscale{0.8}
\plotone{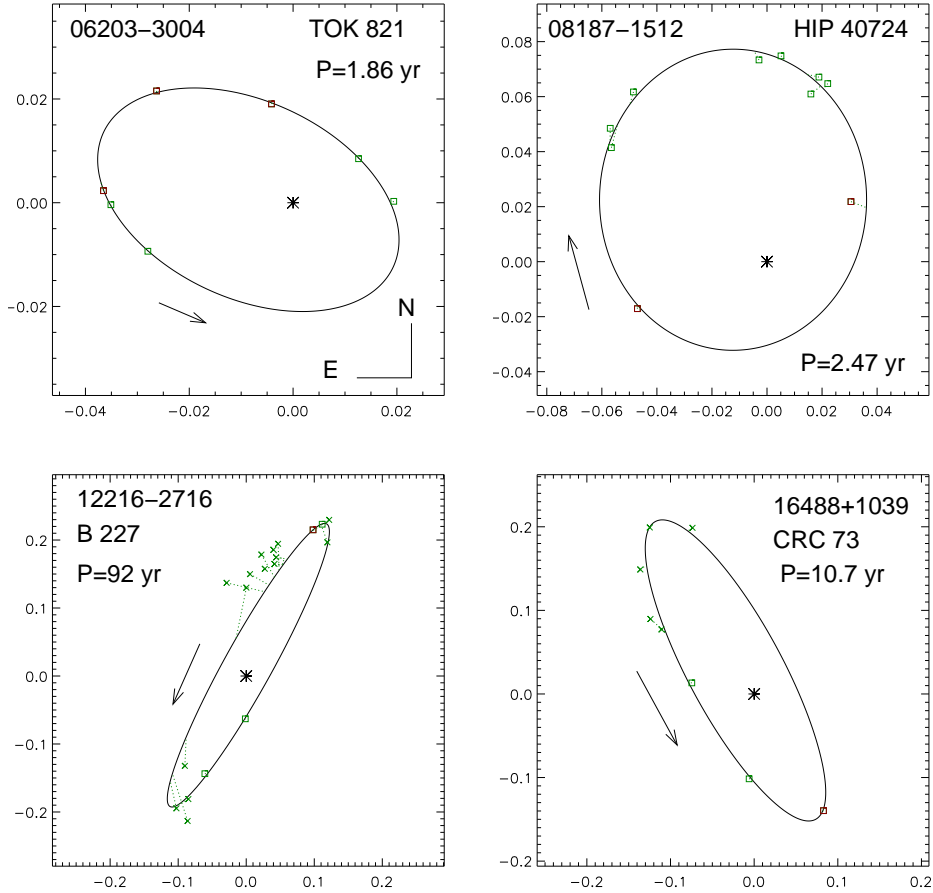}
%\plotone{Orbits.eps}
\caption{Four orbits computed here for the first time but, nevertheless,
  relatively well constrained. See the caption to Figure~\ref{fig:calorb}.
\label{fig:orb} }
\end{figure*}

Figure~\ref{fig:orb} illustrates  some orbits determined here  for the
first time.  The bright  ($V=3.0$ mag) star  $\zeta$~CMa (HIP  30122, HR
2282,  06203$-$3005) has  been  resolved  at SOAR  in  2019.95 and  is
designated as TOK~821 in the WDS.  A spectroscopic orbit with a period
of 675  days has been  published by \citet{Colacevic1941}.  Fixing the
period to its  spectroscopic value, the remaining  visual elements are
fitted to the 7 measurements made  at SOAR so far. Using the Hipparcos
parallax of 9.0$\pm$0.13 mas, the orbit gives a mass sum of 12.2 \msun
for this pair of spectral type B2.5V. 
%[EDR3 plx 9.49mas, RUWE=5.2].

The K5V  dwarf HIP 40724 (08187$-$1512)  has been resolved at  SOAR in
2018.25 into a binary with  a substantial magnitude difference $\Delta
I \approx 2.5$ mag.  Its orbit with  $P = 2.5$ yr uses the unpublished
measure   in  2016.04   communicated   by   T.~Henry  (2021,   private
communication). The  pair passed through  the periastron in  2021, and
its measurements at close separations are tentative. With the parallax
of 28.85$\pm$0.56  mas, the orbit  corresponds to  the mass sum  of 1.4
\msun.  The  classical visual  binary B 227  (12216$-$2716, HIP~60281,
F5V) was  discovered by \citet{B1928} in  1926.  Its 92 yr  orbit uses
the  micrometer  measurements   (with  suitable  quadrant  adjustment,
considering     the    small     $\Delta     m$),    two     published
speckle-interferometric   measurements   in   the   1980s,   and   two
measurements made at SOAR in 2017 and 2021. Finally, the faint pair of
M-type   dwarfs   CRC   73   (16488+1039)  discovered   in   2014   by
\citet{CrC2017} and observed  at SOAR three times in  2018--2021 has a
quasi-circular orbit with $P = 10.7  \pm 1.2$ yr. Despite the grade 4,
the  orbit is  well  constrained,  and the  errors of  its elements  are
modest.

%-------------------------------------------------------------
\subsection{Spurious Pairs}
\label{sec:spurious}

\startlongtable

\begin{deluxetable}{ l l l l  } 
\tabletypesize{\scriptsize}    
\tablecaption{Likely Spurious Pairs
\label{tab:bogus} }                   
\tablewidth{0pt}     
\tablehead{ \colhead{WDS}  &
\colhead{Discoverer}  &  
\colhead{Resolved} & 
\colhead{Unresolved\tablenotemark{a}} 
}
\startdata  % plx cmt P*
00077$-$5615 & HDS 15 & 0\farcs1 HIP       & 2021, L, R  \\ % RUWE=1.0 plx 0.93mas 0.142 P*=481
00259$-$3112 & HDS 58 & 0\farcs2 HIP       & 2021, L, R   \\ % RUWE=1.3 0.91mas
00292$-$3755 & HDS 65 & 0\farcs2 HIP       & 2021, L, R  \\ % RUWE 0.8 plx 0.76
00374$-$3904 & WHI 1  & 0\farcs3 Sp 1987  & 2021, R  \\ % RUWE=1.2 4.6
00386$-$3903 & WHI 2  & 0\farcs8 Sp 1987   & 2021, L, R   \\ % RUWE 1.0, 0.86
00537$-$7910 & HDS 116 & 0\farcs1 HIP      & 2021, L, R  \\ % RUWE 1.1 0.82  
00546$-$8240 & HDS 120 & 0\farcs2 HIP      & 2021, L, R \\ % R  1.8 1.24
01475$-$0753 & HDS 240 & 0\farcs1 HIP      & 2021, L, R \\ % RUWE 1.1  1.52
01511$-$7832 & HDS 251 & 0\farcs2 HIP      & 2021, L, R  \\ % RUWE 1.5 1.2
01519$-$3120 & HDS 254 & 0\farcs2 HIP      & 2021, L, R \\ % RUWE 1.0 0.85 
01571$-$5031 & VOU 20  & 0\farcs4 Vis 1929 & 2021, L, R  \\ % 1.1  3.30
02088$-$6126 & HDS 286 & 0\farcs1 HIP      & 2021,  R   \\  %1.0 2.30 Phys. 11.79''
02109$-$7558 & HDS 293 & 0\farcs3 HIP      & 2014-21   \\ % 2.0 4.04 giant
02376$-$3659 & B 678AB & 0\farcs7 Vis 1927 & 2021  \\ % R (A,C)=1.0,7.7, plx 5.2mas TRIP
03023$-$7154 & FIN 360 & 0\farcs1 Vis 1961 & 2021, R  \\ %  1.0 6.50 
03031$-$2339 & DAM 1296 & 0\farcs1 2015    & 2021  \\ % 0.9 10.2
03047$-$3410 & WHI 4    & 0\farcs2 Sp 1987 & 2021, L, R  \\ % 1.5  1.62
03096$-$0100 & HDS 405  & 0\farcs3 HIP     & 2021, L, R \\ % 1.0 1.54
04112+1538 & CHR 202  & 0\farcs1 Sp 1985-86 & 2018-21, R  \\ % 1.0 5.48 
04118$-$2444 & DAM 312 Aa,Ab & 0\farcs1    & 2021, R  \\ % 1.1 3.37 AB phys.
04129$-$1107 & HDS 534 & 0\farcs2 HIP      & 2021, L, R  \\ % 1.0 1.09
04416$-$4302 & WHI 6   & 0\farcs4 Sp 1987  & 2021  \\ % 1.9 5.15
04513$-$6804 & FIN 362 & 0\farcs1 Vis 1961 & 2010-21 \\ % 0.9 plx 19.92mas
04562$-$0331 & HDS 640 & 0\farcs2 HIP      & 2021, L, R  \\ % 1.0 0.82
05093$-$3813 & HDS 677 & 0\farcs2 HIP      & 2021, L, R  \\ % 1.1 0.54
05343$-$3626 & HDS 737 & 0\farcs1 HIP      & 2021,L, R  \\ % 1.2 0.76
05397$-$0042 & BAL 677 Aa,Ab & 1\farcs0 2003 & 2021, L, R  \\ % 1.1 0.81
05488$-$2507 & B 90   & 0\farcs2 Vis 1925  & 2021, R  \\ % plx 8.87     
06113$-$1635 & HDS 844 & 0\farcs1 HIP      & 2021, L \\ % 2.1 1.39 
06125$-$3515 & HDS 849 & 0\farcs2 HIP      & 2021, L, R  \\ % 1.1 1.57
06226$-$6238 & HDS 872 & 0\farcs1 HIP      & 2021, L, R  \\ % 1.1 1.12
06358$-$1028 & HDS 901 & 0\farcs2 HIP      & 2021, R   \\ % 1.3 2.56   
06447$-$3213 & PRO 27  & 1\farcs1 Vis 1911  & 2021, L \\ % 2.4 1.23
06596$-$2754 & SEE 73  & 0\farcs3 Vis 1897  & 2021, R  \\ % 1.0 6.3
07374$-$3330 & HDS 1079 & 0\farcs5 HIP       & 2021, L, R  \\ % 1.2 0.87
08227$-$1555 & HDS 1194 & 0\farcs2 HIP       & 2021, L, R  \\ % 0.9 0.69  
08542$-$6142 & RST 4901 & 0\farcs3 Vis 1942-69 & 2018-21, L,R \\ % 0.9 1.23 
15219$-$3445 & HDS 2162 &  0\farcs1 HIP      & 2021, L, R  \\ % 1.1 1.93
15596$-$5612 & HDS 2251  & 0\farcs1 HIP      & 2021, L, R  \\ % 0.8, 0.96
16015$-$5416 & HDS 2258 & 0\farcs2 HIP      & 2021, L, R  \\ % 1.5 0.75 
16048$-$6902 & HDS 2271 & 0\farcs2 HIP      & 2021, L, R  \\ % 1.1, 0.76
16118$-$5532 & HDS 2290 & 0\farcs2 HIP      & 2021, L, R  \\ % 0.8 0.96
16243$-$5921 & HDS 2317Aa,Ab & 0\farcs3 HIP & 2021, L, R \\ % AC at 0.91'' is new
16398$-$4706 & HDS 2363 & 0\farcs1 HIP      & 2018-21, L, R  \\ % 1.0 0.70
17001$-$1639 & HDS 2406 & 0\farcs2 HIP      & 2021, R  \\ % 1.0 2.25
17026$-$5504 & HDS 2411 Aa,Ab & 0\farcs2 HIP & 2021, L, R  \\ % 0.9, 0.71
17105$-$1343 & HDS 2425 & 0\farcs2 HIP      & 2021, R  \\ % 1.0 2.70
17118$-$2626 & HDS 2429 & 0\farcs2 HIP      & 2021, L, R  \\ % 0.8 1.00
17238$-$2138 & HDS 2456 & 0\farcs1 HIP      & 2021, R  \\ % 0.9 2.39
17253$-$5600 & HDS 2459 & 0\farcs2 HIP      & 2021, L, R  \\ % 1.5 1.60
17254$-$1643 & HDS 2461 & 0\farcs2 HIP      & 2017-21, L, R  \\ % 1.3 0.50
17317$-$0959 & HDS 2474 & 0\farcs1 HIP      & 2021, R  \\ % 0.9 3.00
17419$-$0944 & HDS 2503 & 0\farcs1 HIP      & 2021, R  \\ % 0.8 2.61 
17482$-$2801 & HDS 2514 & 0\farcs1 HIP      & 2021, L, R  \\ % 0.9 0.30
17492$-$3315 & HDS 2515 & 0\farcs2 HIP      & 2021, L, R  \\ % 0.7 0.70
17582$-$1916 & CHR 66   & 0\farcs4 Sp 1983  & 2021, L, R  \\ % 1.1 3.08
17583$-$5010 & HDS 2533 & 0\farcs2 HIP      & 2021, L  \\ % 2.3 1.03    
17587$-$1152 & HDS 2535 & 0\farcs1 HIP      & 2021, R  \\ % 1.1, 2.27
18052$-$3921 & HDS 2549 & 0\farcs3 HIP      & 2021, L, R  \\ % 0.8 0.41
18174$-$2730 & HDS 2585 & 0\farcs2 HIP      & 2021, L, R  \\ % 0.7 2.19
18197$-$4542 & CHR 148  & 0\farcs3 Sp 1989  & 2008-21  \\ % 24.8 astrom 0.32y 
18218$-$5526 & HDS 2597 & 0\farcs1 HIP      & 2021, L, R  \\ % 0.9 1.14
18272$+$0012 & STF2316Aa,Ab & 0\farcs1 vis 1951 & 2009-21 \\ % 2.2 3.3
18320$-$0607 & HDS 2629 & 0\farcs1 HIP      & 2021, L, R  \\ % 1.0 0.75
18351$-$1659 & MCA 52 & 0\farcs2 Sp 1980    & 2017-21, L, R  \\ % 0.7 0.61
18387$-$1429 & HDS 2641 AB & 0\farcs1 HIP   & 2014-21 \\ % Wrong coord? G=18.5??  
18423$-$0720 & HDS 2649 & 0\farcs1 HIP      & 2008-21, R  \\ % 0.8 2.75
18501$-$0823 & HDS 2670 & 0\farcs1 HIP      & 2013-21, L, R  \\ % 0.9 0.49 3*Neg
19020$-$1705 & OCC 9061 & 0\farcs1 Occ 2014 & 2021, R  \\ % 0.9 7.63
19024$-$2541 & HDS 2699 & 0\farcs2 HIP      & 2021, R  \\ % 1.3 4.11
19044$-$0541 & HDS 2704 & 0\farcs2 HIP      & 2021, R  \\ % 0.8 2.33
19205$-$0525 & ISO 10 Aa,Ab & 0\farcs5 Sp 1987 & 2021  \\ % 3.3  22.69 real? Orb 0.73yr
19223$-$2226 & I 1399 & 0\farcs4 Vis 1925   & 2021 \\ % 3.8 2.16 real?
19230$-$0144 & HDS 2743 & 0\farcs1 HIP      & 2021, L, R  \\ % 0.9 1.06
19293$-$6833 & HDS 2769 & 0\farcs2 HIP      & 2021, L, R  \\ % 1.0 1.34
19317$-$3153 & VOU 85 & 0\farcs1 Vis 1937   & 2021, L, R  \\ % 1.0 1.35
19383$-$1527 & ARU 16 & 0\farcs7 Sp 1984    & 2021, L, R  \\ % 0.9 2.20
19420$-$2223 & HDS 2793 & 0\farcs1 HIP      & 2021, L, R  \\ % 0.9 1.00
19425$-$1607 & BU 1288 & 0\farcs2 Vis 1889  & 2021, R  \\ % 0.7 18.36 
19460$-$0735 & HDS 2808 & 0\farcs1 HIP      & 2021, L, R  \\ % 1.1 1.69
19492+0817 & ARU 17   & 0\farcs4 Sp 1984  & 2021, R  \\ % 1.0 6.11
20007$-$1757 & ARU 18  & 0\farcs4 Sp 1984   & 2021, L, R  \\ % 0.9 2.00
20020$-$1754 & ARU 19  & 0\farcs4 Sp 1984   & 2021, L, R  \\ % 1.0 1.63
20029$-$8455 & HDS 2859 & 0\farcs2 HIP      & 2021, L, R  \\ % 1.0 2.46
20109$-$0637 & ARU 20   & 0\farcs5 Sp 1984  & 2021, L, R  \\ % 1.7 $-$1.13
21104$-$4114 & ARU 3 & 0\farcs2  Sp 1984    & 2021, R  \\ % 0.9 7.90
21415$-$7723 & BLM 6 & 0\farcs1 Sp 1976     & 2021 \\ % V=3.77 UR HIP
21420$-$2940 & HDS 3090 &  0\farcs1 HIP     & 2021, L, R  \\ % 1.1 1.73
22456$-$8529 & JSP 836 & 0\farcs7 Vis 1929  & 2021, R  \\ %1.1 13.81
22461$-$1210 & ARU 5   & 0\farcs3 Sp 1982   & 2014-21 \\ % 12.2 6.84
22473$-$3410 & CHR 189 AB & 0\farcs3 Sp 1992 & 2015-21, R  \\ % 1.1 9.02
22532$-$1102 & HDS 3249 & 0\farcs2 HIP      & 2021,  L, R  \\ % 1.0 2.33
23114$-$4259 & B 594   & 0\farcs2 vis 1925-63 & 2008-21   \\ % 0.9 1.7mas Orb 21yr Nrr1983!! N=11 WDS
23256$-$2556 & HDS 3332 & 0\farcs2 HIP      & 2021,  L, R  \\ % 1.1 0.86
\enddata
\tablenotetext{a}{Additional indications of the spurious nature of visual pairs:
R -- no excess noise in Gaia EDR3, RUWE$<$2; 
L -- long estimated period; 
%S -- short estimated period or spectroscopic coverage;   
%Vib -- artefact caused by telescope vibration. 
}
\end{deluxetable}
%02336$-$3724 & HDS 334 & 0\farcs2 HIP      & 2021  \\  % 2.9 2.78 real?
%06389$-$3250 & HDS 918 & 0\farcs1 HIP      & 2021  \\ % 2.8 2.25  
%07012$-$2557 & LSC 136 & ??                 & 2021  \\ % 13.0 67.8   
%08470$-$4518 & HDS 1271 & 0\farcs2 HIP       & 2021 \\ % 2.1 2.03 
%11557$-$1438 & RST 3761 & 0\farcs1 Vis 1937-1989 & 2018-2021 \\ % 2.3  4.47 real
%18237$+$2146 & TOK 60Aa,Ab & 0\farcs04 Sp 2009 & 2018-21  \\ % 26.96mas 2.3 astrometric?
%18271$-$0853 & HDS 2608 & 0\farcs2 HIP      & 2021  \\ 6.1 3.69 
%18408$-$1320 & HDS 2646 & 0\farcs1 HIP      & 2021 \\ % 5.6 3.53 real? 
%18457$-$4226 & HDS 2660 & 0\farcs1 HIP      & 2018-2021 \\ % 3.5 3.20 

In 2021,  we observed a  substantial number of ``neglected''  pairs that
have  only  one  documented   resolution  and  no  further  confirming
observations.   Our  measurements  provide such  confirmation  and  in
several     cases      reveal     additional      close     subsystems
(section~\ref{sec:new}). Some pairs were  not resolved because orbital
motion decreased their separation below  the detection limit of HRCam.
Such binaries are candidates  for future observations and, eventually,
orbit calculation.   However, the majority of  unresolved double stars
are spurious discoveries.   A list of spurious pairs  was presented in
\citep{SAM18}; here  we continue  this effort and  list   another 94
apparently  spurious  pairs  in  Table~\ref{tab:bogus}.   Its  columns
contain WDS  code, discoverer  designation, separation  in arcseconds,
observing method  (HIP ---  Hipparcos, Sp ---  speckle interferometry,
Vis ---  visual micrometer, Occ  --- occultation), and years  when the
pair was resolved.  The last column  gives the years when the pair was
not resolved  at SOAR  (some targets  have multiple  observations) and
additional indications  from Gaia that  the star is  single, explained
below. Pairs in Table~\ref{tab:bogus} have  an `X' code added to their
entry in the WDS, indicating the pair is not real.

Gaia EDR3  does not  contain measurements of  close double  stars, but
gives several clues concerning  their reality.  Pairs with separations
from 0\farcs15  to 0\farcs7  and modest  magnitude difference  have no
parallaxes and PMs in EDR3.  Gaia astrometric solutions with excessive
noise caused by  companions are distinguished by the  reduced 
unit weight error (RUWE) parameter, which is less than 1.4 for point sources
and can be larger for binaries. Furthermore, the Gaia parallaxes allow
us to evaluate  orbital periods from projected  separations.  Take for
example the  first entry  in Table~\ref{tab:bogus},  00077$-$5615 (HDS
15). Hipparcos  measured in 1991.25  a separation of  0\farcs142 that
corresponds to 153 au, given the parallax of 0.93\,mas from Gaia EDR3.
The orbital semimajor axis cannot be less than half of the separation,
so the  minimum period computed  by the third  Kepler's law is  481 yr
(assuming a mass sum of 2 \msun). The real period is likely longer, so
the hypothesis  that this pair  has closed down  in 30 yr  between the
Hipparcos  and  SOAR observations  cannot  be  true.  HDS~15  is  thus
definitely  spurious,  and  RUWE=1.0  confirms this.   The  letters  L
(long-period)   and   R  (small   RUWE)   in   the  last   column   of
Table~\ref{tab:bogus}  mark  similar  cases. An  opposite  example  is
11557$-$1438 (RST  3761).  This pair has  been consistently unresolved
at SOAR between 2018 and 2021.  However, it was measured several times
between 1937  and 1989, has  a RUWE=2.3, and  the parallax of  4.5 mas
suggests an orbital period  on the order of 200 yr. So,  RST 3761 is a
real binary,  it just closed down  at present. In our  future work, we
will check  neglected pairs using  Gaia to avoid wasting  telescope time
for their observation.

Most  entries  in  Table~\ref{tab:bogus}   are  Hipparcos  pairs  with
separations of 0\farcs2  or less. This is below  the diffraction limit
of the 30 cm Hipparcos aperture. The majority of Hipparcos pairs
with  comparable separations  are  real, and a  small  number of  spurious
resolutions   could    be   caused    by   problems   in    the   data
reduction. Spurious binaries discovered  by visual observers and never
confirmed are  also found  in our  list. The  third group  of spurious
pairs results  from speckle interferometry. Several  false discoveries
have been made by HRCam  and later retracted \citep[see][]{SAM18}. The
discovery codes  ARU \citep{Argue1985}, WHI \citep{White1991},  and ISO
\citep{Isobe1991} correspond to early  speckle programs that produced
a number spurious pairs owing to lack of experience with this technique.

%-------------------------------------------------------------
%\subsection{}

%-------------------------------------------------------------
\section{Summary and outlook}
\label{sec:sum}

The total  number of  observations made  with HRCam  to date  is about
30,300.  This paper documents the  observations made in 2021 and their
use for improving the orbits. It  reports 50 newly resolved pairs (the
majority of which are members  of triple and higher-order systems) and
gives a list of likely spurious pairs checked at SOAR.

Speckle interferometry was invented half a century ago and now it
has completely  replaced the visual micrometer  measurements. The time
span of speckle  data becomes comparable to the 100--200  yr of visual
coverage, and in some cases the  latter loses its scientific value, as
illustrated   by   the   first   orbit   in   Figure~\ref{fig:calorb}.
Furthermore, the historic visual double stars, still a majority of the
WDS entries, are complemented by  the new Hipparcos and speckle pairs.
Millions  of  Gaia  pairs  will  soon  completely  dominate  over  the
present-day  sample of  double stars.   So, will  the current  speckle
programs  be viable  in the  future, or  will they  progressively lose
their importance,  retaining only  a historic  value, like  the visual
micrometer measurements or the Ptolemeus astrometric catalog?   We
  believe that in the near term,  the demand for new speckle data will
  actually increase, being driven by Gaia and other missions.

The  Gaia  catalog  already   contains  accurate  relative  positions,
motions, and parallaxes  of many thousands of double  stars wider than
$\sim$0\farcs7 and gives hints on the closer pairs (missing parallaxes
or a large RUWE). Accurate parallaxes  add value to the visual orbits,
enabling  good measurement  of stellar  masses.  Although  the current
Gaia parallaxes of visual binaries are often biased by orbital motion,
this caveat  will be corrected in  the next data release  3 which will
account for acceleration.  Unbiased parallaxes of single components in
visual triples, available now in the EDR3, can also serve for the mass
measurent of their inner orbital pairs.

The huge  future impact  of Gaia  on the binary  star research  can be
gleaned  by looking  back  at the  Hipparcos  mission, which  revealed
thousands  of  close pairs  missed  by  the historic  visual  surveys.
Follow-up speckle monitoring of  the Hipparcos binaries (including the
work reported  here) leads  to the determination  of their  orbits and
masses. The Gaia binaries will  require a similar follow-up, but their
large number  will necessitate a selective strategy. Determination of
all  potentially accessible  orbits (the  default approach  until now)
will  be replaced  by the  astrophysically motivated  samples. Obvious
candidates are low-mass nearby stars \citep[see e.g.][]{Vrijmoet2022},
high-mass stars, pre-main-sequence stars, binaries hosting exoplanets,
and  hierarchical systems.   So, the  demand for  new observations  of
close visual binaries will likely increase in the near future.

\begin{acknowledgments} 

 We are grateful to the anonymous referee for a thorough check of the data.
We thank the SOAR operators for efficient support of this program, and
the SOAR director  J.~Elias for allocating some  technical time.  This
work is based in part on observations carried out under CNTAC programs
CN2019A-2, CN2019B-13, CN2020A-19, CN2020B-10, and CN2021B-17.  R.A.M.
and  E.C.    acknowledge  support  from  the   FONDECYT/CONICYT  grant
No. 1190038.  The research of A.T. is supported by the NSFs NOIRLab.

This work  used the  SIMBAD service operated  by Centre  des Donn\'ees
Stellaires  (Strasbourg, France),  bibliographic  references from  the
Astrophysics Data  System maintained  by SAO/NASA, and  the Washington
Double Star  Catalog maintained  at USNO.  This  work has made  use of
data   from   the   European   Space   Agency   (ESA)   mission   Gaia
(\url{https://www.cosmos.esa.int/gaia})  processed  by  the  Gaia  Data
Processing      and     Analysis      Consortium      (DPAC,     {\url
  https://www.cosmos.esa.int/web/gaia/dpac/consortium}). Funding for the
DPAC  has been provided  by national  institutions, in  particular the
institutions participating in the Gaia Multilateral Agreement.

\end{acknowledgments} 

\facility{SOAR}.

%\clearpage
%\landscape

%\LongTables
%\input{orbtable1.tex}

\end{document}